\documentclass[%
 reprint,
superscriptaddress,
 amsmath,amssymb,
 aps,
]{revtex4-2}

\usepackage{graphicx}
\usepackage{dcolumn}
\usepackage{bm}
\usepackage[colorlinks=true,urlcolor=blue,citecolor=blue,linkcolor=blue]{hyperref}
\usepackage[mathscr]{euscript}

\begin{document}

\title[Force Measurement]{Simple Method for the Direct Measurement of Cohesive Forces \\ Between Microscopic Particles}

\author{Johnathan Hoggarth}%
 \affiliation{Department of Physics and Astronomy, McMaster University, 1280 Main Street West, Hamilton, L8S 4M1, Ontario, Canada}

\author{Kari Dalnoki-Veress}%
\email{dalnoki@mcmaster.ca}
\affiliation{Department of Physics and Astronomy, McMaster University, 1280 Main Street West, Hamilton, L8S 4M1, Ontario, Canada}%
\affiliation{UMR CNRS Gulliver 7083, ESPCI Paris, PSL Research University, Paris, 75005, France}

\date{\today}

\begin{abstract}
We present a  simple and inexpensive method for measuring weak cohesive interactions. This technique is applied to the specific case of oil droplets with a depletion interaction,  dispersed in an aqueous solution. The experimental setup involves creating a short string of droplets while immobilizing a single droplet. The droplets are held together via depletion interactions and a single cohesive bond holds together nearest neighbours. Initially, the buoyant droplets  are held in a flat horizontal chamber. The droplets float to the top of the chamber and are in contact with a flat glass interface. In the horizontal configuration, there is no component of the effective buoyant force acting in the plane of the chamber. The angle of the chamber is gradually increased and the effective buoyant force acting on the string of droplets slowly increases. At a critical point,  when the combination of gravity and  buoyancy is equal to the cohesive force, the droplet string will detach from the immobile droplet. Our method allows for a simple direct measurement of cohesive forces on the tens of pico-Newton scale. To illustrate the validity of this technique, the droplet radii and concentration of depletant are varied, and their impact on the cohesive force is measured. This method offers a simple, accessible, and reproducible means of exploring cohesive interactions beyond the specific case of oil droplets and a depletion interaction.
\end{abstract}


\maketitle

\section{Introduction}\label{sec1}

Attractive interactions between small particles are common in various materials, including colloidal systems, cellular aggregates, foams, and emulsions. The ability to quantify these interactions is essential for understanding the stability, mechanical properties, and overall behaviour of these materials. One specific example is emulsions, which are ubiquitous in pharmaceutical, cosmetic, and food industries. The stability of an emulsion significantly impacts its physical properties~\cite{bibetteEmulsionsBasicPrinciples1999}. Oil-water emulsions are the most common example and can be stabilized by surfactants, which adsorb at the oil-water interface and decrease the surface tension. Surfactants can also form micelles in solution which act as a depletant leading to attractive interactions between droplets due to depletion forces~\cite{vrijPolymersInterfacesInteractions1976b}. The depletion interaction can be described through the Asakura-Oosawa potential \cite{asakuraInteractionParticlesSuspended1958}. The interactions between the oil droplets are fundamental to the stability and mechanical properties of the emulsion~\cite{jamiesonForcesOilDrops2019}. Quantifying these interactions is essential to fully understand the properties of an emulsion.

Many techniques have been developed to directly measure attractive interactions between small particles, with depletion forces being a common example~\cite{kleshchanokDirectMeasurementsPolymerinduced2008a}. Techniques used to measure depletion forces include the surface forces apparatus~\cite{richettiDirectMeasurementDepletion1992}, optical tweezers~\cite{crockerEntropicAttractionRepulsion1999, nilsen-nygaardStabilityInteractionForces2014, griffithsMeasuringInteractionPair2016, chenDeterminationInteractionMechanism2018, liaoContactlessMeasurementsSaltenhanced2023}, atomic force microscopy~\cite{millingDirectMeasurementDepletion1995, dagastineDynamicForcesTwo2006, taborMeasurementAnalysisForces2012}, micropipette based techniques~\cite{brochard-wyartUnbindingAdhesiveVesicles2003, chuJohnsonKendallRobertsTheoryApplied2005, ono-dit-biotContinuumModelApplied2020b}, and total internal reflection microscopy~\cite{sharmaDirectMeasurementDepletion1996, rudhardtDirectMeasurementDepletion1998}. Many of these techniques require specialized materials and instruments. While force measurements using these techniques can provide precise measurements, the cost and time investment required, make these techniques inaccessible to many laboratories. Being able to quickly measure the attractive forces acting between microscopic particles is beneficial in experiments where cohesion is an important variable.

For slightly deformable cohesive spherical particles, such as oil droplets in an emulsion, the Johnson-Kendall-Roberts (JKR) theory describes the force required to separate two bodies~\cite{johnsonSurfaceEnergyContact1997}. The pull-off force between two spherical vesicles was derived by Brochard and de~Gennes as~\cite{brochard-wyartUnbindingAdhesiveVesicles2003}:
\begin{equation}
    F_\mathcal{A} = \pi R \mathcal{A},
    \label{eqn:Fa}
\end{equation} where $R$ is the radius of the vesicles and $\mathcal{A}$ is the cohesive force per unit length. We will call $\mathcal{A}$ the cohesive strength.

Previously, Ono-dit-Biot \textit{et al.} directly measured the force required to pull apart two cohesive oil droplets in an aqueous solution of sodium dodecyl sulfate (SDS) and sodium chloride (NaCl) using the micropipette deflection technique~\cite{ono-dit-biotContinuumModelApplied2020b}. They found that Eqn.~\ref{eqn:Fa} adequately describes the pull-off force of two droplets, and that, as expected~\cite{vrijPolymersInterfacesInteractions1976b},  $\mathcal{A}$ scales linearly with the micelle concentration. 
While this micropipette deflection technique is capable of precisely measuring the pull-off force for a range of micelle concentrations and droplet radii, the process of measuring the force is labour  intensive.

Here we present a simple experiment to measure the cohesion between small particles like colloids, cells, or droplets. We illustrate the approach with oil droplets in a surfactant solution. This method allows for rapid measurement of weak cohesive forces down to tens of pico-Newtons of force. The protocol presented here provides a simple experiment where the cohesion between particles can be measured across multiple sizes and cohesion strengths. Our method provides results consistent with previous measurements of the depletion force in an identical system using a micropipette deflection technique~\cite{ono-dit-biotContinuumModelApplied2020b}. Furthermore, we use a similar experimental setup with a different geometry to validate our measurements by observing the critical force required to break a cohesive bond in a droplet triad; and validate that friction or pinning of the particles within the chamber does not affect the results.

\section{Methods}\label{sec2}

In the present section, we describe in detail the method used to measure the cohesive forces between oil droplets. Oil droplets are produced in a chamber containing an aqueous surfactant solution and due to buoyancy, they will float to the top of the chamber. When the chamber is held at an angle, the driving force due to buoyancy and gravity acting on the droplets in the plane of the chamber is modified. For simplicity we refer to this as the \emph{effective buoyancy}\footnote{To be precise,  by `effective buoyancy',  we mean the difference between buoyancy and gravity which depends on the difference in density of the components acting parallel to the top surface of the chamber i.e. there is  no component of the effective buoyancy acting on the droplets when the chamber is horizontal.}, $F_b$, and $F_b \propto \sin{\theta}$, where $\theta$ is the angle of the chamber with respect to the horizontal. We use a motorized rotation stage (RV240CC, Newport Corporation, USA) controlled by a motion controller (ESP300, Newport Corporation, USA) to slowly increase the component of the effective buoyancy acting along the plane of the sample chamber. The rotation stage can easily be replaced with a manual stage. A small cluster of droplets is formed in the horizontal configuration. One of the droplets is immobilized while the angle is slowly increased. As the angle is increased, the effective buoyancy increases to the point where a bond breaks, providing the strength of a bond. A critical aspect of the method is immobilizing a particle, so as to measure the bond strength with its neighbour. However, we stress that immobilizing a particle can be achieved in many ways, for example suction using a micropipette to hold a droplet in place~\cite{ono-dit-biotContinuumModelApplied2020b}. The method we present here is simply what was most accessible, and is not a critical aspect of the methodology. Further details regarding the experimental chamber, aqueous solution composition, droplet preparation, and imaging are provided below.

Experimental chambers are constructed by sandwiching a 3D printed spacer between two glass slides. The size of the spacer is 30 x 20 x 5 mm$^3$. The spacer has a   single inlet at the front to allow the chamber to be filled with solution.   The size of the inlet is 5 x 2 mm$^2$ which is large enough to allow for air to flow out of the chamber while it is being filled with solution.   To ensure minimal contact forces between oil droplets and the top glass slide, the slides are pre-cleaned in an aqueous bath of Sparkleen soap solution (Fisher Scientific, Canada) using an ultrasonic bath for 5 minutes.    For the force measurements described in Section \ref{sec:force_measurements}, the top glass slide is   modified using photolithography. A 100 $\mu$m film of photoresist SU-8 (Kayaka Advanced Materials, USA) is deposited on the top glass slide and patterned into a 2D hopper with a narrow opening smaller than the diameter of a single droplet. The opening is large enough for a portion of a droplet to stick out while still being immobilized. The size of the hopper can be adjusted such that a wide range of droplet sizes can be probed.
\begin{figure}
    \centering
    \includegraphics[width=\linewidth]{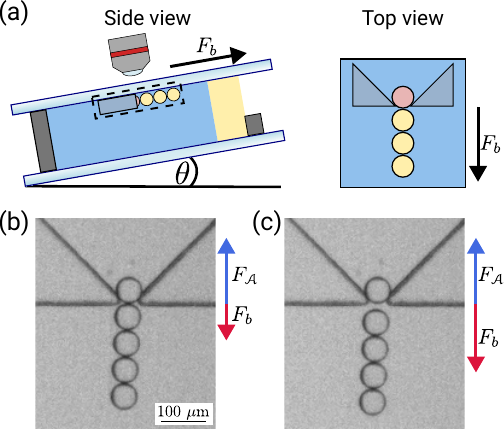}
    \caption{\textbf{(a)} Side view and top-down view schematic diagram of the experimental chamber.   The chamber is filled with an aqueous surfactant solution (blue) and sealed with mineral oil (yellow) to prevent evaporation. A single droplet is held stationary while a chain of droplets is dangling from it. The chain sticks to the stationary droplet due to depletion interactions. The effective buoyant force pulling the chain away from the stationary droplet increases with the angle $\theta$ of the chamber.   (b) Microscopy image of a chain of droplets attached to a stationary droplet.   (c) The chain detaches from the stationary droplet once the effective buoyant force is greater than the cohesive force.}
    \label{fig:1}
\end{figure}

An aqueous solution of SDS (Sigma-Aldrich, USA) and NaCl (Caledon Laboratories, Canada) is added to the chamber. The concentration of SDS is varied across different experiments while the concentration of NaCl is held constant at 1.5 \% (w/w). The SDS serves to stabilize the oil droplets against coalescence. In addition, SDS is a micelle forming molecule and provides attraction between particles through the depletion interaction. In pure water, the critical micelle concentration (CMC) of SDS is 8.3 mM. However, in the presence of 1.5 \% NaCl, the ionic interactions between the SDS molecules are screened and the formation of micelles is more energetically favourable, decreasing the CMC to 1 mM~\cite{thevenotAggregationNumberCritical2005c, naskarCounterionEffectMicellization2013b}. The concentration of micelles in solution is given by $C_m = C_\textrm{SDS} - \textrm{CMC}$, where $C_\textrm{SDS}$ is the concentration of SDS in solution. Previously, it was found that the cohesive force between droplets scales linearly with the concentration of micelles~\cite{ono-dit-biotContinuumModelApplied2020b, bibetteCreamingEmulsionsRole1992b}. The density of the surfactant solution was calculated for each concentration of SDS ranging from $\rho = 1015~$kg/m$^3$ for   $C_m$ = 0 mM  to $\rho = 1070~$kg/m$^3$ for   $C_m$ = 184 mM.   After filling the chamber with the SDS solution, a layer of light paraffin oil (Supelco, MilliporeSigma, USA) was added on top to prevent evaporation and ensure the concentration of SDS remains constant throughout the experiments.

Near mono-disperse oil droplets are produced using the snap-off instability~\cite{barkleySnapproductionMonodisperseDroplets2015a,barkleyPredictingSizeDroplets2016c}. Droplets are made by flowing light paraffin oil ($\rho_\textrm{oil} = 840~$kg/m$^3$) through a glass micropipette with a tip radius of 10 $\mu$m to 30 $\mu$m. The micropipette is connected to a translation stage and the position of the droplets being produced can be controlled. When droplets are produced, the rotation stage is initially held horizontally. A single oil droplet is deposited on one side of the hopper opening,   such that a small portion of the droplet sticks out of the hopper but is otherwise held stationary by the walls of the hopper. A string of droplets is then produced on the other side of the opening, such that a single droplet is touching the stationary droplet. Since the force acting on the immobilized droplet depends on the number of droplets to which that droplet is attached, the number of droplets in the string can be adjusted to accommodate the force required to break the cohesive bond for each experiment. A schematic of the droplets is shown in Fig.~\ref{fig:1}(a).   Images of the experiment are captured from above using optical microscopy, allowing individual droplets to be resolved. A camera (DCC1240M, Thorlabs, USA) is used with a 2x objective lens (Edmund Optics, USA). Illumination is provided from below using an edge-lit collimated LED backlight (Advanced Illumination, USA), ensuring uniform lighting of the chamber. A representative microscopy image from an experiment is shown in Fig.~\ref{fig:1}(b). The angle of the chamber is slowly increased at a rate of 0.01~degrees/s   while the frame rate is set to 0.5 frames per second. Images are taken every 0.02 degrees. As the angle of the chamber is increased, the effective buoyant force acting on the string of droplets is increased. At a critical angle, the string of droplets will break off of the stationary droplet as shown in Fig.~\ref{fig:1}(c). At this point, the pull off force is equal to the effective buoyant force.

\section{Force Measurements}
\label{sec:force_measurements}

  Before measuring the cohesive forces between droplets, we first quantified the force of friction between the oil droplets and the top glass slide to determine if friction is significant. If this force is similar to the adhesion, then it must be included in the force balance. To quantify the frictional force, a single droplet was deposited while the chamber was horizontal. The angle of the chamber was slowly increased until the droplet began moving across the top glass slide. The angle of the chamber was then decreased to the point where the droplet would move in the opposite direction. Half the difference between these angles then determines the upper bound of the effect of friction between the droplet and the top glass slide (note that this also aids in determining precisely the horizontal position). For a droplet of size $R = (43.7\pm0.3)~\mu$m, the angle at which the droplet moved was $0.100 \pm 0.025$ degrees, resulting in a contribution of friction less than $1.1\pm0.3$~pN. Since the forces we measure are hundreds of pico-Newtons, the frictional force is neglected. Furthermore, since our measured forces were taken when the angle of the chamber was well above 0.1 degrees, this further negates the impact of frictional forces because as the angle increases, the normal force acting on the droplet decreases since $N \propto \cos{\theta}$ and the frictional force $F_f \propto N$. 

At the moment when the string of droplets gets detached from the stationary droplet, the magnitude of the cohesive force is equal to that resulting from the effective buoyancy:
\begin{equation}
    F_\mathcal{A} = F_b = N \frac{4}{3} \pi R^3 \Delta \rho  g \sin{\theta},
    \label{eqn:1}
\end{equation}
where $N$ is the number of droplets in the string (excluding the immobilized droplet), $\Delta \rho$ is the difference in density between the oil droplets and surfactant solution, $g$ is the acceleration due to gravity, and $\theta$ is the angle of the experimental chamber. For a given radius of droplet, the effective buoyant force acting on the bond between the immobilized droplet and its neighbour can be controlled by either changing the angle of the chamber or by changing the number of droplets attached to the string. This allows us to probe a wide range of forces. Even though $\theta$ ranges from 0$^\circ$ - 90$^\circ$, $N$ can be set arbitrarily high to investigate stronger forces.

\begin{figure}
    \centering
    \includegraphics[width=\linewidth]{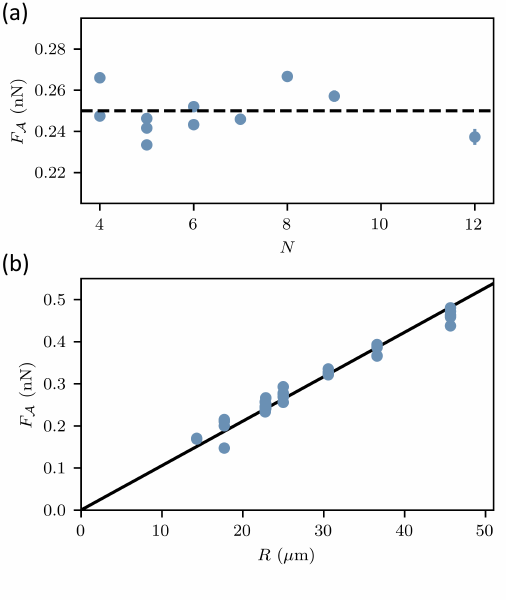}
    \caption{\textbf{(a)} Pull off force $F_\mathcal{A}$ plotted as a function of the number of droplets dangling from the immobile droplet. The average pull off force 0.25 $\pm 0.01$ nN is shown as the dashed line. \textbf{(b)} Pull off force $F_\mathcal{A}$ plotted against droplet radii at a constant concentration of micelles $C_m =71$ mM. The equation of the line of best fit is given by   $F_\mathcal{A} = (3.36 \pm 0.03)\pi R$~pN}.
    \label{fig:2}
\end{figure}

We first confirm that the number of droplets in the string does not impact the measured force. For constant radius and concentration of micelles, the force required to break a bond between droplets should remain constant, regardless of how many droplets are present in the droplet string. We observe that the force required to pull apart a droplet remains constant ($F_b = 0.25 \pm 0.01$ nN) within the resolution of the experiment, as shown in Fig.~\ref{fig:2}(a), while changing the number of droplets in the dangling string from 4 to 12. There is no observable trend when changing $N$.

We next investigated the pull off force as a function of the droplet radius.  Fig.~\ref{fig:2}(b) shows the cohesive force as a function of droplet radii for constant $C_m = 71$ mM. The measurements are consistent with Eqn \ref{eqn:Fa} with an equation of the line of best fit   $F_\mathcal{A} = (3.36 \pm 0.03)  \pi R$~pN.

\begin{figure}
    \centering
    \includegraphics[width=\linewidth]{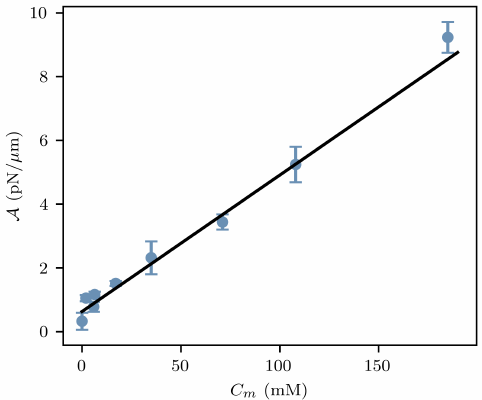}
    \caption{Cohesive strength $\mathcal{A}$ plotted against the concentration of micelles in solution. As the concentration of micelles increases, the cohesive strength increases linearly. Each data point is plotted as a mean with standard deviations. The equation for the line of best fit is given by $\mathcal{A} = [(0.0428 \pm 0.0007) \textrm{pN}/(\mu \textrm{m} \, \textrm{mM})]C_m + (0.63 \pm 0.04) \textrm{pN}/\mu \textrm{m}$.}
    \label{fig:3}
\end{figure}

Next, we investigate the cohesive strength across a range of SDS concentrations. Eqn \ref{eqn:Fa} can be rearranged to solve for the cohesive strength:
\begin{equation}
    \mathcal{A} = \frac{F_\mathcal{A}}{\pi R}.
\end{equation}When increasing $C_m$, the cohesive strength between droplets  increases as shown in prior studies~\cite{ono-dit-biotContinuumModelApplied2020b}. We plot $\mathcal{A}$ as a function of $C_m$ in Fig.~\ref{fig:3} and find, as expected, a linearly increasing trend. We note that there is a non-zero force required to pull apart the droplets even in the absence of micelles. Even though the depletion force is then absent, there are still forces present due to the combination of van der Waals attraction and double layer repulsion, which can be described through Derjaguin–Landau-Verwey–Overbeek theory~\cite{chenDeterminationInteractionMechanism2018}. The y-intercept of the best fit represents these forces.

\section{Further Validation}
\label{sec:validation}

\begin{figure}
    \centering
    \includegraphics[width=\linewidth]{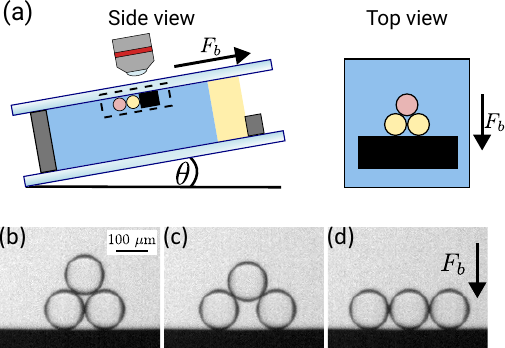}
    \caption{\textbf{(a)}   Side view and top-down view schematic diagram of the experimental chamber used for additional validation experiments. \textbf{(b)-(d)} Collapse of a stable triad of droplets at at a boundary while the effective buoyant force is steadily increased by increasing the angle of the chamber. Representative images \textbf{(b)} before, \textbf{(c)} during, and \textbf{(d)} after the collapse.}
    \label{fig:4}
\end{figure}

Since the experimental technique allows us to quickly measure and modify the cohesive and effective buoyant forces, we are capable of   testing other geometries to further validate our force measurements.   For example, consider a triad of droplets arranged as a triangle which is stable at a boundary [see Fig.~\ref{fig:4}(a)]. Here, the boundary is a cleaved Si wafer which provides a sharp edge~\cite{hoggarthTwodimensionalSpreadingFrictionless2023b}.   A schematic diagram of the chamber is shown in Fig.~\ref{fig:4}(a). We can increase the chamber angle to increase the effective buoyancy and observe the force required to break the stable arrangement of droplets. The pile will collapse once the effective buoyancy is greater than cohesive bond joining  the two droplets at the boundary. A simple force balance   based on this geometry shows that the triad is stable as long as $F_b \leq 4 F_\mathcal{A}$.   Here, $F_b$ is the effective buoyant force acting on the top droplet. We performed experiments with $R=63.4\pm 0.4~\mu$m  and $C_m = 34$~mM as shown in Fig.~\ref{fig:4}. At a critical angle $\theta_c$, where $F_b = 4F_\mathcal{A}$, we have:
\begin{equation}
   \left( \frac{\mathcal{A}}{\Delta \rho g\sin{\theta_c}}\right) \left(\frac{3}{R^2} \right) = 1.
    \label{eqn:Balance}
\end{equation}
We find that with 12 repeated experiments, we obtain a value for the left-hand side of Eqn.~\ref{eqn:Balance} of  $1.08 \pm 0.08$, which is in agreement with what is expected.

\begin{figure}
    \centering
    \includegraphics[width=\linewidth]{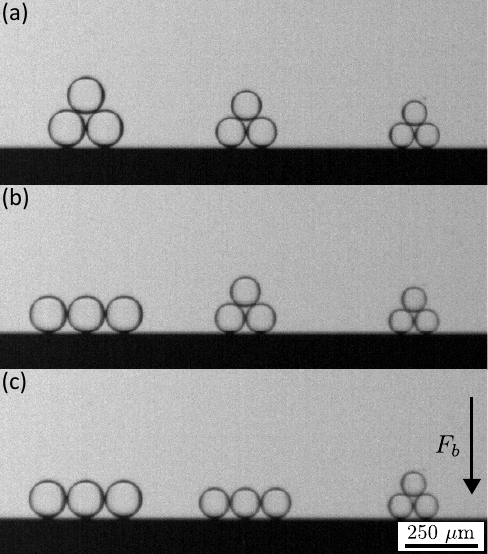}
    \caption{Piles of droplets of varying sizes collapsing as the effective bouyancy increases. As the angle increases, going from (a) to (c), the piles of droplets becomes unstable and collapses.}
    \label{fig:5}
\end{figure}

Lastly, we show an example of three triads with three different sizes of droplets [$R=(63.4\pm0.4)~\mu$m, $(52.6\pm0.3)~\mu$m, and  $(43.7\pm0.3)~\mu$m] as shown in Fig.~\ref{fig:5}. We observe that as the angle is increased and the effective buoyant force increases, the largest triad collapses first, followed by the medium sized droplet triad, while the triad prepared from the smallest droplets remains stable against collapse. This is to be expected from Eqn.~\ref{eqn:Balance}, since we see that $\sin \theta_c \propto 1/R^2$, keeping all other parameters constant.   These supplemental experiments validated the measurements taken in Section \ref{sec:force_measurements}. However we note that there are potentially additional frictional forces between the oil droplets and the Si wafer, which makes these experiments less precise than the previous measurements. Simply, these experiments validate that the cohesive forces measured previously align with the force required to collapse a triad. Lastly, we note that the square-root of the first factor of the product in Eqn.~\ref{eqn:Balance} is a length-scale, much like a granular analog of the gravity-capillary length used to describe a liquid meniscus, and has been used to describe the spreading of cohesive oil droplets in two and three dimensions~\cite{ono-dit-biotContinuumModelApplied2020b, hoggarthTwodimensionalSpreadingFrictionless2023b}.

\section{Conclusions}\label{sec4}

Within this paper we share a simple experimental protocol capable of quickly measuring the cohesive pull off force between particles. The technique is particularly useful since it provides the ability to measure the cohesive forces quickly across a range of particle sizes without the need for specialized instruments and materials. The method presented is simple and accessible to a wide range of laboratories. Although this method was validated in this study by measuring the depletion interaction between oil droplets, it is applicable to a wide range of scenarios. The only two requirements to measure the force is a way to immobilize a single buoyant particle and a way to modify the effective buoyant force by tilting the chamber. 

  While our technique provides a quick and simple method for measuring the cohesive forces between microscopic particles, we acknowledge that our approach cannot be extended to all cases. First and foremost, the force of friction or contact forces between the particles and the top surface must be negligible, or well characterized, such that the force balance used to calculate cohesion is accurate. We have outlined a method to test the impact of friction by observing the hysteresis of droplet motion at small angles and showed that for our emulsion system, friction was negligible. Additionally, our method requires that the particles are stable and do not coalesce, requiring a surfactant present in solution. As well, our technique requires that the particles can be observed over time as the cohesive bond breaks. For our system, we utilized simple optical microscopy, which sets a limit to the difference in the index of refraction between the oil and water phase of the emulsion. However, the precise method to observe particles is not critical and other imaging techniques, like fluorescence microscopy, can be used. Overall, the method described within this paper offers a flexible technique with potential applications for a broad range of particles.

\subsubsection*{Author Contribution Statement}
JH was responsible for the design, execution and analysis of the experiments with support from KDV. JH wrote the first draft of the manuscript and KDV edited the manuscript.

\subsubsection*{Data Availability Statement} 
The datasets generated during the current study are available from the corresponding author upon reasonable request.

\bibliography{References.bib}

\end{document}